\documentclass[twocolumn,aps,superscriptaddress]{revtex4}

\usepackage{amssymb}
\usepackage{amsmath}
\usepackage{mathrsfs}
\usepackage{graphicx}
\usepackage{subfigure}
\usepackage[normalem]{ulem}
\usepackage[dvips]{color}

\renewcommand{\sout}{\bgroup \color{red} \ULdepth=-.5ex \ULset}

\begin{document}

\title{Isospin splitting of pion elliptic flow in relativistic heavy-ion collisions}
\author{He Liu\footnote{liuhe@sinap.ac.cn}}
\affiliation{Shanghai Institute of Applied Physics, Chinese Academy of Sciences, Shanghai 201800, China}
\affiliation{University of Chinese Academy of Sciences, Beijing 100049, China}
\affiliation{Cyclotron Institute and Department of Physics and Astronomy, Texas A$\&$M University, College Station, TX77843-3366, USA}
\author{Feng-Tao Wang\footnote{wangfengtao@sinap.ac.cn}}
\affiliation{Shanghai Institute of Applied Physics, Chinese Academy of Sciences, Shanghai 201800, China}
\affiliation{University of Chinese Academy of Sciences, Beijing 100049, China}
\author{Kai-Jia Sun\footnote{kjsun@tamu.edu}}
\affiliation{Cyclotron Institute and Department of Physics and Astronomy, Texas A$\&$M University, College Station, TX77843-3366, USA}
\author{Jun Xu\footnote{corresponding author: xujun@zjlab.org.cn}}
\affiliation{Shanghai Advanced Research Institute, Chinese Academy of Sciences, Shanghai 201210, China}
\affiliation{Shanghai Institute of Applied Physics, Chinese Academy of Sciences, Shanghai 201800, China}
\author{Che Ming Ko\footnote{ko@comp.tamu.edu}}
\affiliation{Cyclotron Institute and Department of Physics and Astronomy, Texas A$\&$M University, College Station, TX77843-3366, USA}

\date{\today}

\begin{abstract}
Based on the framework of an extended multiphase transport model with mean-field potentials in both the partonic phase and the hadronic phase, we explain the elliptic flow difference between $\pi^+$ and $\pi^-$ in the beam-energy scan program at the relativistic heavy-ion collider by incorporating the vector-isovector potential for quarks and antiquarks with different isospins. It is found that the isospin splitting of pion elliptic flow favors a strong vector-isovector interaction, and thus serves as a probe of the quark matter equation of state as well as the QCD phase structure at finite baryon and isospin chemical potentials.
\end{abstract}

\maketitle

Understanding the properties of the strongly interacting quark-gluon plasma (QGP) as well as the hadron-quark phase transition is one of the main goals of relativistic heavy-ion collision experiments. The elliptic flow ($v_2$) characterizing the different numbers of particles moving in the reaction plane and out of the reaction plane is a good probe of the QGP properties~\cite{Pos98,Vol08}. The number-of-constituent-quark (NCQ) scaling of $v_2$~\cite{STAR04,STAR05a,STAR05b,PHENIX07a,PHENIX07b,STAR08} in ultrarelativistic heavy-ion collisions serves as an evidence of the existence of the QGP, indicating that the elliptic flow is mostly developed in the partonic phase. Recently, 'low-energy' relativistic heavy-ion collisions were carried out in the beam-energy scan (BES) program at the relativistic heavy-ion collider (RHIC), in order to search for the signal of the critical point of the hadron-quark phase transition. Among various interesting phenomena different from those observed in ultrarelativistic heavy-ion collisions is the splitting of $v_2$ between protons and antiprotons, $K^+$ and $K^-$, and $\pi^+$ and $\pi^-$~\cite{STAR13}, showing the breakdown of the NCQ scaling law. The observed $v_2$ splitting can be due to the larger $v_2$ for transported quarks than that for produced quarks, or similarly, their different rapidity dependencies~\cite{Dun11,Gre12,Iva13}, hydrodynamic evolution of the QGP at finite baryon chemical potential~\cite{Ste12,Hat15}, and the smaller radial flow of particles than their antiparticles~\cite{Sun15}. Although the $v_2$ splitting between $\pi^-$ and $\pi^+$ as well as its linear dependence on the charge asymmetry was proposed to be attributed to the electric quadrupole moment in the produced QGP induced by the chiral magnetic wave~\cite{Bur11}, it is disfavored by transport simulations based on the chiral kinetic equations of motion~\cite{Sun16,Han19}. In our previous study~\cite{Xu14,Xu16}, we have shown that the $v_2$ splitting between protons and antiprotons as well as between $K^+$ and $K^-$ can be attributed to the different mean-field potentials for particles and their antiparticles. We will demonstrated in the present study that the $v_2$ splitting between $\pi^+$ and $\pi^-$ can be explained by the different mean-field potentials for particles with different isospins.

The elliptic flow in heavy-ion collisions at intermediate energies was reguarded as a probe of the mean-field potential or the nuclear matter equation of state~\cite{Dan02}. In noncentral heavy-ion collisions at relativistic energies, the produced QGP is of an almond shape in the transverse plane. Particles with attractive potentials are more likely to be trapped in the system and move in the direction perpendicular to the reaction plane, while those with repulsive potentials are more likely to leave the system and move along the reaction plane, thus reducing and enhancing their respective elliptic flows. In the baryon-rich matter produced in low-energy relativistic heavy-ion collisions, light quarks and baryons are affected by a more repulsive potential compared to light antiquarks and antibaryons as a result of the vector coupling, leading to the $v_2$ splitting between protons and antiprotons as well as between $K^+$ and $K^-$~\cite{Xu14,Xu16}. Since heavy-ions are neutron-rich nuclei, the produced matter is not only baryon-rich but also neutron-rich or $d$-quark-rich, where $d$ $(\bar{u})$ quarks are expected to have a more repulsive potential than $u$ $(\bar{d})$ in the presence of a vector-isovector interaction. This will enhance the $v_2$ of $\pi^-$ while reduce the $v_2$ of $\pi^+$, qualitatively consistent with that observed experimentally. Since the stopping power becomes larger with the decreasing collision energy, the medium at midrapidities is expected to be not only more baryon-rich but also more neutron-rich, again qualitatively consistent with the larger $v_2$ splitting between $\pi^+$ and $\pi^-$ at lower collision energies. The isospin splitting of charged pion $v_2$ can thus be used to constrain the strength of the vector-isovector interaction and to help determine the quark matter equation of state as well as the QCD phase structure at finite baryon and isospin chemical potentials~\cite{Liu16}.

The present study is based on the framework of an extended multiphase transport model (extended AMPT)~\cite{Xu14,Xu16,Guo18}. In this model the momentum distribution of initial partons is from melting hadrons generated by the heavy-ion jet interaction generator (HIJING) model~\cite{Xnw91} through the Lund string fragmentation model with its parameters fitted to particle multiplicities at RHIC-BES energies~\cite{Zha16}. The coordinates of these partons in the transverse plane are set to be the same as those of the colliding nucleons from which they are produced, while their coordinates in the longitudinal direction are uniformly distributed within the finite thickness of the participant matter by considering the Lorentz contraction. The evolution of the partonic phase is then described by a relativistic Boltzmann transport approach, with the mean-field potentials of partons from a 3-flavor Nambu-Jona-Lasinio (NJL) model, and isotropic two-body scatterings with the cross section fitted by the final charged-particle elliptic flow. The mean-field potentials depend on the local phase-space distribution functions of partons, which are calculated numerically from averaging over parallel events using the test-particle method~\cite{Won82,Ber88}. The partonic phase ends when the central energy density is below 0.8 GeV/fm$^3$, and then a coalescence model is used to describe the hadronization procedure. Different from the naive spatial coalescence algorithm for hadronization, we now use an improved coalescence algorithm~\cite{Wan19} by selecting the combination of a quark-antiquark pair, three quarks, or three antiquarks, which has the largest Wigner function rather than the closest distance in coordinate space. In the improved coalescence scenario, the baryon, charge, and strangeness numbers for a single event are conserved, while mixed-event combinations are allowed for choosing partons close in phase space from different events in the spirit of the test-particle method, so that the parton combinations can have a large Wigner function. The improved hadronization algorithm prefers combination in not only coordinate space but also momentum space, thus preserves better the flow of partons during the hadronization procedure. The species of hadrons are determined by the flavor of its constituent quarks as well as their invariant mass~\cite{Lin05}. The evolution of these hadrons is described by a relativistic transport (ART) model~\cite{Li95}, with the mean-field potentials for pions, kaons, and baryons as well as antibaryons incorporated~\cite{Xu12}, in addition to various elastic collisions, inelastic collisions, and decay channels satisfying the detailed balance condition. The isovector hadronic mean-field potentials relevant for the present study are the pion $s$-wave potential and the symmetry potential for baryons with different isospins. The pion $s$-wave potential, which is repulsive for $\pi^-$ and attractive for $\pi^+$ in neutron-rich medium, is taken from Ref.~\cite{Kai01}. The pion $p$-wave potential has shown to be less important compared to the $s$-wave potential~\cite{Xu10,Xu13,Zha17}. The momentum-independent symmetry potential incorporated in the hadronic phase leads to a more repulsive potential for neutron-like baryons, e.g., $\Delta^-$, and a more attractive potential for proton-like baryons, e.g., $\Delta^{++}$, in the neutron-rich medium. The effects of these hadronic isovector potentials are, however, expected to be largely suppressed due to the low baryon density of the hadronic phase, originating from a more realistic finite thickness of initial partonic medium and the hadronization criterion employed in the present study compared with those used in Ref.~\cite{Xu12}.

The mean-field potentials of partons are given by the 3-flavor NJL model with isovector interactions, and its Lagrangian can be written as~\footnote{The signs for vector interaction terms are set to be negative to be consistent with the relativistic mean-field model, different from those in Ref.~\cite{Liu16}.}
\begin{eqnarray}
\mathcal{L}_{\rm NJL} &=& \bar{q}(i\rlap{\slash}\partial-\hat{m})q
+\frac{G_S}{2}\sum_{a=0}^{8}[(\bar{q}\lambda_aq)^2+(\bar{q}i\gamma_5\lambda_aq)^2]
\notag\\
&-&\frac{G_V}{2}\sum_{a=0}^{8}[(\bar{q}\gamma_\mu\lambda_aq)^2+
(\bar{q}\gamma_5\gamma_\mu\lambda_aq)^2]
\notag\\
&-&K\{\det[\bar{q}(1+\gamma_5)q]+\det[\bar{q}(1-\gamma_5)q]\}
\notag\\
&+&G_{IS}\sum_{a=1}^{3}[(\bar{q}\lambda_aq)^2+(\bar{q}i\gamma_5\lambda_aq)^2]\notag\\
&-&G_{IV}\sum_{a=1}^{3}[(\bar{q}\gamma_\mu\lambda_aq)^2+(\bar{q}\gamma_5\gamma_\mu\lambda_aq)^2].
\end{eqnarray}
In the above, $q = (u, d, s)^T$ and $\hat{m} = \text{diag}(m_u,m_d,m_s)$ are the quark fields and the current quark mass matrix for $u$, $d$, and $s$ quarks, respectively; $\lambda_a$ are the Gell-Mann matrices with $\lambda_0$ = $\sqrt{2/3}I$ in the 3-flavor space with the SU(3) symmetry; $G_S$ and $G_V$ are respectively the scalar-isoscalar and the vector-isoscalar coupling constant; and the $K$ term represents the six-point Kobayashi-Maskawa-t'Hooft interaction that breaks the axial $U(1)_A$ symmetry~\cite{Hoo76}. The additional $G_{IS}$ and $G_{IV}$ terms represent the scalar-isovector and the vector-isovector interactions, with $G_{IS}$ and $G_{IV}$ the corresponding coupling constants, respectively. Since the Gell-Mann matrices with $a = 1,2,3$ are identical to the Pauli matrices in $u$ and $d$ space, the isovector couplings break the SU(3) symmetry while keeping the isospin symmetry. In the present study, we employ the parameters $m_u = m_d = 3.6$ MeV, $m_s = 87$ MeV, $G_S\Lambda^2 = 3.6$, $K\Lambda^5 = 8.9$, and the cutoff value in the momentum integral $\Lambda = 750$ MeV given in Refs.~\cite{Bra12,Lut92}. As is known, the position of the critical point for the chiral phase transition is sensitive to $G_V$~\cite{Asa89,Fuk08,Bra12}, which was later constrained within $0.5G_S<G_V<1.1G_S$ from the relative $v_2$ splitting between protons and antiprotons as well as between $K^+$ and $K^-$ in relativistic heavy-ion collisions~\cite{Xu14}. In the present study, we choose $G_V=1.1G_S$ throughout the calculation.

In the mean-field approximation and considering only the flavor-singlet state for the vector-isoscalar term, the single-particle Hamiltonian for partons with flavor $i$ ($i=u,d,s$) can be written as
\begin{eqnarray}\label{eq2}
H_i &=& \sqrt{M_i^2+{p_i^{\ast}}^2}\pm \frac{2}{3}G_V(\rho^0_u+\rho^0_d+\rho^0_s)\nonumber\\
&&\pm G_{IV}\tau_{3i}(\rho^0_u-\rho^0_d).
\end{eqnarray}
Here we take the convention that the upper (lower) sign is for quarks (antiquarks). $M_i$ is the constituent quark mass determined by the gap equation
\begin{eqnarray}
M_i = m_i-2G_s\sigma_i+2K\sigma_j\sigma_k-2G_{IS}\tau_{3i}(\sigma_u-\sigma_d),
\end{eqnarray}
where $\sigma_i = \langle q_i \bar{q_i}\rangle$ is the quark condensate, ($i$, $j$, $k$) is any permutation of ($u$, $d$, $s$), and $\tau_{3i}$ is the isospin quantum number of quarks, i.e., $\tau_{3u} =1$, $\tau_{3d} = -1$, and $\tau_{3s} = 0$. $\vec{p}^{\ast}_i = \vec{p}_i \mp \frac{2}{3}G_V(\vec{\rho_u}+\vec{\rho_d}+\vec{\rho_s}) \mp G_{IV} \tau_{3i} (\vec{\rho}_u - \vec{\rho}_d )$ is the real momentum of partons, with $\vec{p}_i$ being the canonical momentum and $\vec{\rho}_i \equiv \langle \bar{q_i} \vec{\gamma} q_i \rangle$ being the space components of the net quark density. $\rho^0_i \equiv \langle \bar{q_i}\gamma^0 q_i \rangle$ in Eq.~\eqref{eq2} is the time component of the net quark density. The quark condensate and the 4-component net quark density can be calculated from the phase-space distributions of quarks ($f_i$) and antiquarks ($\bar{f_i}$) through the relations
\begin{eqnarray}
\langle \bar{q_i} q_i \rangle &=& -2N_c\int_0^\Lambda\frac{d^3p}{(2\pi)^3}\frac{M_i}{\sqrt{M_i^2+p_i^2}}(1-f_i-\bar{f}_i),\\
\langle \bar{q_i} {\gamma}^{\mu} q_i \rangle &=& 2N_c\int_0^\Lambda\frac{d^3p}{(2\pi)^3} \frac{p^{\mu}}{\sqrt{M_i^2+p_i^2}}(f_i-\bar{f_i}),
\end{eqnarray}
where $2N_c$ is the quark spin and color degeneracy. In the transport approach, the whole system is divided into grids with $f_i$ and $\bar{f_i}$ at each cell calculated from averaging over parallel events using the test-particle method. In this way, the spatial distributions of the above quantities can be obtained.

Given the single-particle Hamiltonian as Eq.~\eqref{eq2}, partons with flavor $i$ evolve according to the following canonical equations
\begin{eqnarray}
\frac{d\vec{r}_i}{dt} &=& \frac{\partial H_i}{\partial \vec{p}_i} = \frac{\vec{p}^{\ast}_i}{\sqrt{M_i^2+{p_i^{\ast}}^2}}, \\
\frac{d\vec{p}^{\ast}_i}{dt} &=& -\frac{\partial H_i}{\partial \vec{r}_i} \mp \frac{d}{dt} [\frac{2}{3}G_V(\vec{\rho}_u+\vec{\rho}_d+\vec{\rho}_s)\nonumber\\
&&+ G_{IV}\tau_{3i}(\vec{\rho}_u-\vec{\rho}_d)].
\end{eqnarray}
The time component of the vector-isovector potential $G_{IV}\tau_{3i}(\rho^0_u-\rho^0_d)$ is expected to give a more repulsive (attractive) potential for $d$ ($u$) quarks in the baryon-rich and $d$-quark-rich partonic phase. The space component of the vector-isovector potential $-G_{IV} \tau_{3i} (\vec{\rho}_u - \vec{\rho}_d )$ may have an opposite effect, but its effect is small at the early stage when the net quark flux has not been developed. A similar argument for the vector-isoscalar potential can be found in Ref{s.~\cite{Son12,Ko14}. For the ease of discussion, we define $R_{IV} = G_{IV}/G_{S}$ as the reduced strength of the vector-isovector coupling and will discuss its effects on the isospin splitting of the elliptic flow. The scalar-isovector interaction leads to the constituent mass splitting of $u$ and $d$ quarks, and the effect is only pronounced around the boundary of the chiral phase transition~\cite{Liu16} at the later stage of the partonic phase when the density is rather low, and it has been found to have negligible effects on the isospin splitting of the elliptic flow.


\begin{figure}[tbh]
\centerline{
\includegraphics[scale=0.35]{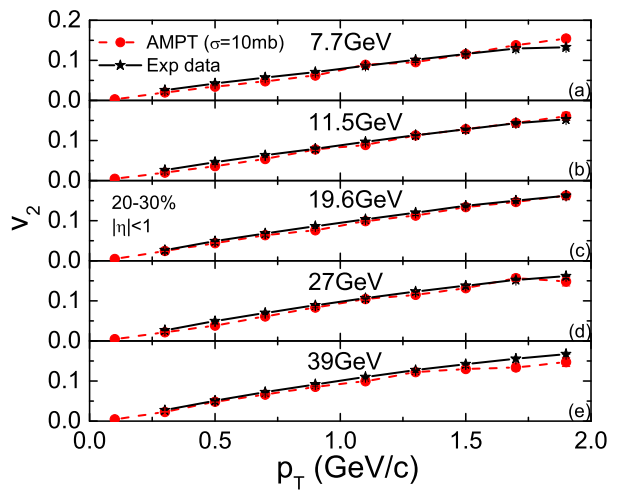}}
\caption{(color online) Comparison of the transverse momentum dependence of charged particle elliptic flows at mid-pseudorapidities ($|\eta|<1$) in mid-central ($20-30\%$) Au+Au collisions at RHIC-BES energies from the extended AMPT model with those measured by the STAR Collaboration~\cite{STAR12}.}
\label{fig1}
\end{figure}

We first fit the parton scattering cross section in order to reproduce the charged particle $v_2$ at RHIC-BES energies. As shown in Fig.~\ref{fig1}, using the same event-plane method~\cite{Pos98,Vol08} as in the experimental analysis, the extended AMPT model with an isotropic parton scattering cross section of 10 mb can reproduce reasonably well the transverse-momentum ($p_T$) dependence of mid-pseudorapidity charged particle $v_2$ at all RHIC-BES energies. The cross section is larger than that in Ref.~\cite{Xu16}, as a result of using an improved coalescence approach. As discussed in Ref.~\cite{Wan19}, a larger cross section is needed to reproduce the same hadron $v_2$ once combinations of partons close in momentum space is favored, compared with the naive spatial coalescence scenario. We have also compared the $p_T$ spectra of mid-rapidity charged pions in central Au+Au collisions at RHIC-BES energies from AMPT calculations with the STAR data in Fig.~\ref{fig1.5}. It is seen that these $p_T$ spectra are reproduced reasonably well, except that the AMPT results give slightly stiffer $p_T$ spectra at lower energies while smaller multiplicities at higher energies. The agreement of the $p_T$ dependence of $v_2$ as well as the $p_T$ spectra serves as a baseline for the following study on the $v_2$ splitting of charged pions.

\begin{figure}[tbh]
\centerline{
\includegraphics[scale=0.3]{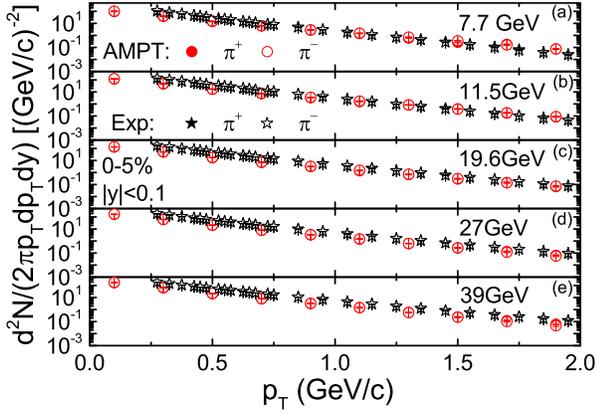}}
\caption{(color online) Comparison of transverse momentum spectra of mid-rapidity ($|y|<0.1$) $\pi^+$ and $\pi^-$ in central ($0-5\%$) Au+Au collisions at RHIC-BES energies from the extended AMPT model with those measured by the STAR Collaboration~\cite{STAR17}.}
\label{fig1.5}
\end{figure}

\begin{figure}[tbh]
\centerline{
\includegraphics[scale=0.3]{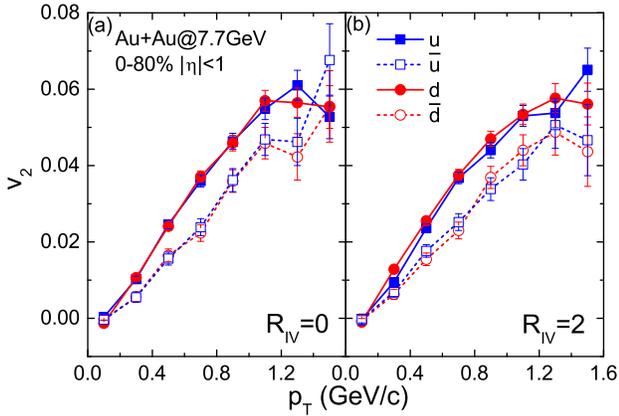}}
\caption{(color online) Transverse momentum dependence of the elliptic flows for mid-pseudorapidity ($|\eta|<1$) light quarks and antiquarks in minibias Au+Au collisions at $\sqrt{s_{NN}} = 7.7$ GeV from the extended AMPT model with (right) and without (left) the vector-isovector interaction.}
\label{v2-q}
\end{figure}

We choose minibias Au+Au collisions at $\sqrt{s_{NN}} = 7.7$ GeV as a representative system to discuss the effect of the vector-isovector interaction on the isospin splitting of the elliptic flow. The transverse-momentum dependence of $v_2$ for light quarks and antiquarks with different isospins are displayed in Fig.~\ref{v2-q}. The $v_2$ splitting between quarks and antiquarks is due to the vector-isoscalar potential. Without the vector-isovector interaction ($R_{IV}=0$), $u$ and $d$ quarks as well as $\bar{u}$ and $\bar{d}$ have the same elliptic flow. With the vector-isovector interaction ($R_{IV}=2$), $d$ quarks have a slightly larger $v_2$ than $u$ quarks. This shows that the time component of the vector-isovector potential has the dominate effect over the space component, leading to a more repulsive (attractive) potential for $d$ ($u$) quarks, as discussed above. Also, $\bar{u}$ ($\bar{d}$) are affected by a more repulsive (attractive) potential according to Eq.~\eqref{eq2}, and have thus a slightly larger (smaller) $v_2$ in the dominating low-$p_T$ region. We note that the increasing parton elliptic flows with transverse momentum are due to parton scatterings as the mean-field potentials mainly lead to a splitting of their elliptic flows.

\begin{figure}[tbh]
\centerline{
\includegraphics[scale=0.3]{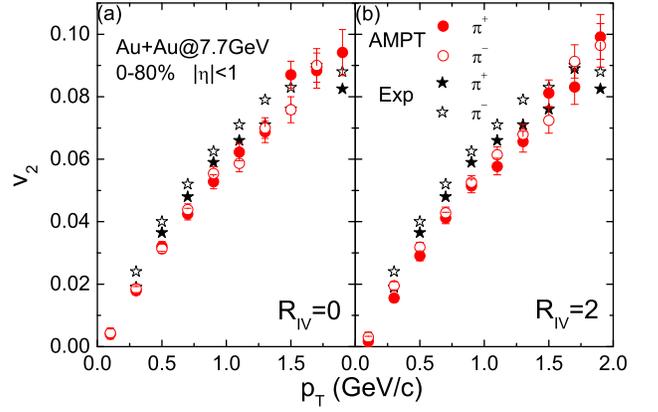}}
\caption{(color online) Transverse momentum dependence of the elliptic flows for mid-pseudorapidity ($|\eta| < 1$) $\pi^+$ and $\pi^-$ in minibias Au+Au collisions at $\sqrt{s_{NN}} = 7.7$ GeV from the extended AMPT model with (right) and without (left) the vector-isovector interaction. The experimental results measured by the STAR Collaboration~\cite{STAR13a} are also plotted for reference.}
\label{v2-h}
\end{figure}

Since the isospin splitting of light quark $v_2$ is preserved during hadronization, a larger splitting of charged pion $v_2$ is obtained as $\pi^+$ and $\pi^-$ are composed of $u\bar d$ and $\bar u d$, respectively. For resonance states with different isospins, e.g., $\rho^{+(-)}$ or $\Delta^{++(-)}$, formed from the quark coalescence, they eventually decay into charged pions and carry information of the isospin-dependent potentials on their constituent light quarks. The final elliptic flows of mid-pseudorapidity $\pi^+$ and $\pi^-$ are displayed in Fig.~\ref{v2-h} for the same collision system as in Fig.~\ref{v2-q}. It is seen that the resulting $v_2$ as a function of transverse momentum from the extended AMPT model has a similar overall magnitude compared with those measured by the STAR Collaboration. A slightly larger $v_2$ for $\pi^-$ than $\pi^+$ is observed with the vector-isovector interaction, while the $v_2$ for $\pi^-$ and $\pi^+$ are similar without the vector-isovector interaction.

\begin{figure}[tbh]
\includegraphics[scale=0.3]{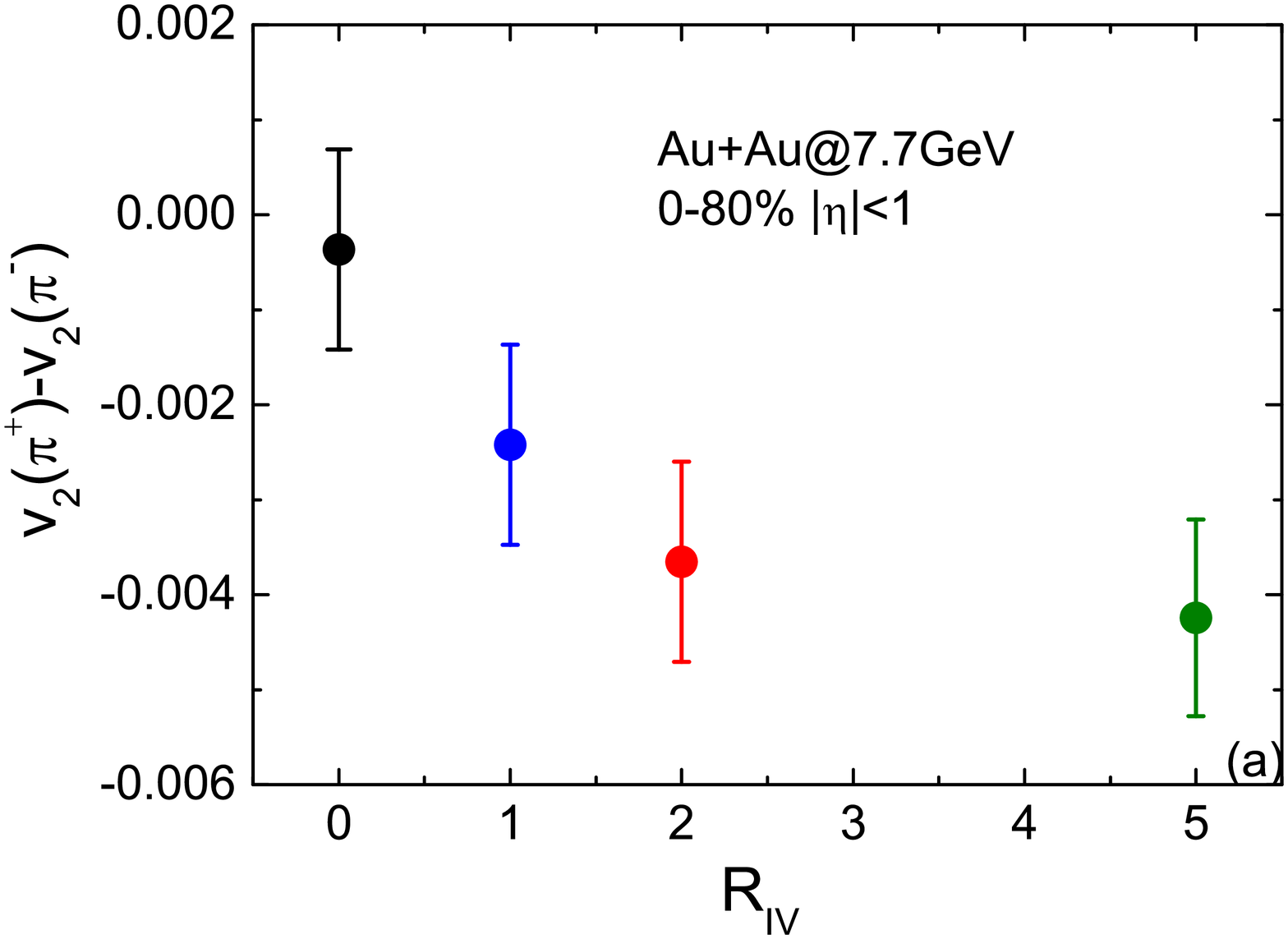}
\includegraphics[scale=0.3]{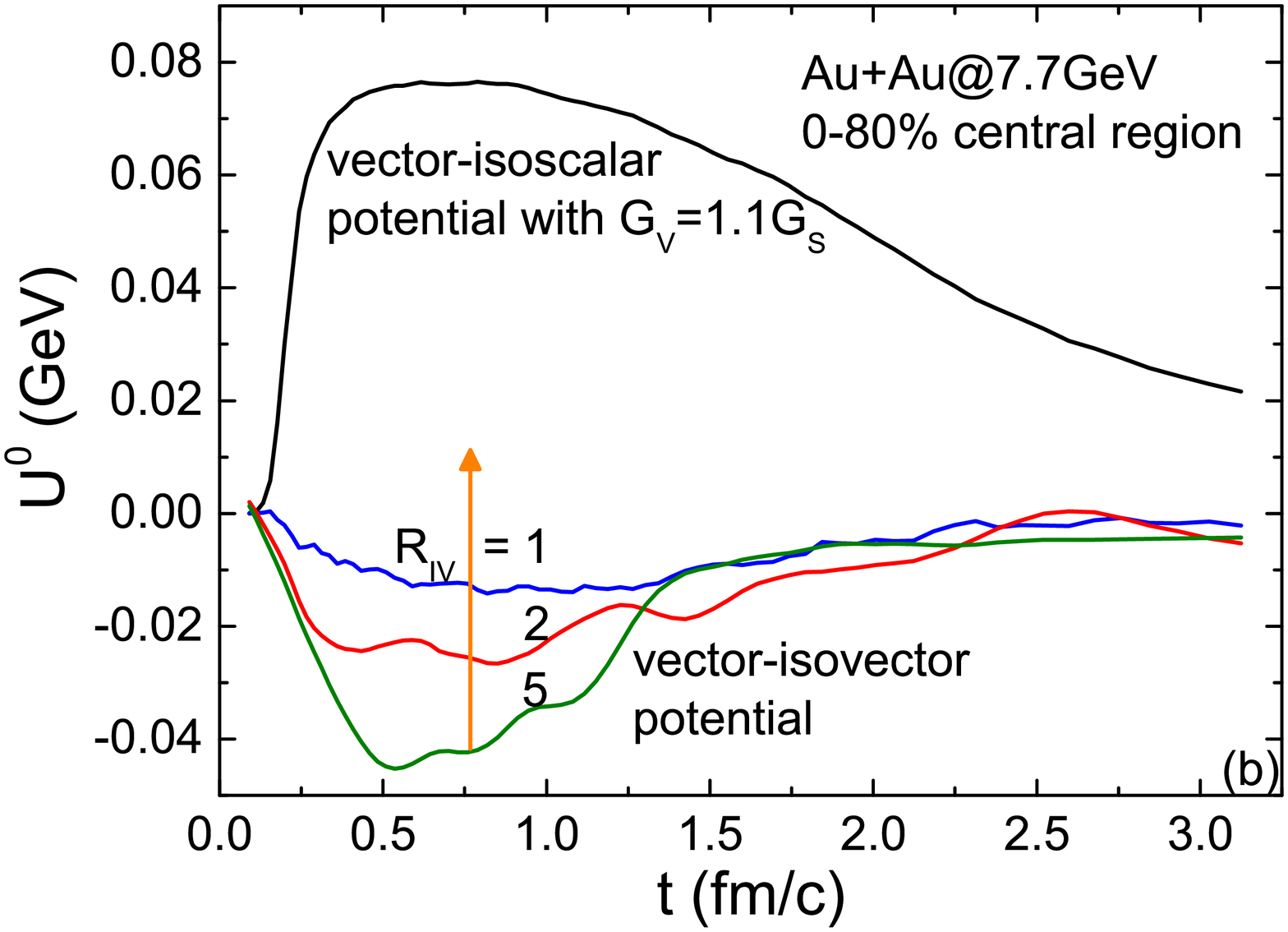}
\caption{(color online) Upper: Elliptic flow difference between mid-pseudorapidity ($|\eta|<1$) $\pi^+$ and $\pi^-$ with different vector-isovector coupling strengths; Lower: Evolution of the time component of the vector-isoscalar potential as well as the vector-isovector potential with different coupling strengths in the central region of the partonic phase. Results are for minibias Au+Au collisions at $\sqrt{s_{NN}}$ = 7.7 GeV from the extended AMPT model.}
\label{v2-U-giv}
\end{figure}

To illustrate more clearly the effect of the vector-isovector interaction, we display in the upper panel of Fig.~\ref{v2-U-giv} the $p_T$-integrated elliptic flow difference between mid-pseudorapidity $\pi^+$ and $\pi^-$ obtained
with different vector-isovector coupling strengths. Without the vector-isovector interaction, although the mean value of the $v_2$ difference is negative due to the weak hadronic potential, it is consistent with 0 within the statistical error. The $v_2$ splitting increases with the increasing coupling strength of the vector-isovector interaction until it is about to saturate for $R_{IV}>2$. To understand the saturation effect from the vector-isovector interaction, we further show in the lower panel of Fig.~\ref{v2-U-giv} the evolution of the time component of the vector-isoscalar potential $U^0_V = \frac{2}{3}G_V(\rho^0_u+\rho^0_d+\rho^0_s)$ as well as the vector-isovector potential $U^0_{IV}=G_{IV}(\rho^0_u-\rho^0_d)$ for different coupling strengths in the central region of the partonic phase. Although the peak value of $U^0_{IV}$ increases with increasing $R_{IV}$, it decreases more rapidly for $R_{IV}=5$ compared with that for $R_{IV}=2$. This is due to the quicker reduction of the isospin asymmetry in the high-density phase as a result of the stronger vector-isovector potential. For the vector-isoscalar potential, it has a much larger magnitude compared to the vector-isovector potential. The saturation effect of the vector-isovector potential is also observed at higher collision energies, where the lifetimes of both potentials become even shorter as a result of a thinner thickness and more rapid expansion of the produced partonic matter.

\begin{figure}[tbh]
\centerline{
\includegraphics[scale=0.35]{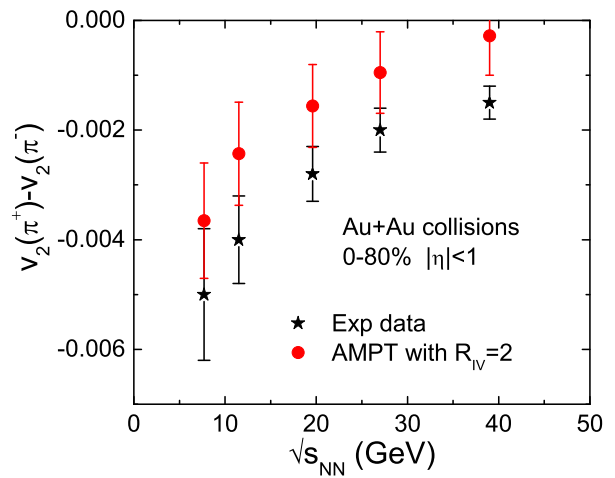}}
\caption{(color online) Comparison of the elliptic flow difference between mid-pseudorapidity ($|\eta|<1$) $\pi^+$ and $\pi^-$ at different RHIC-BES energies from the extended AMPT model with those measured by the STAR Collaboration~\cite{STAR13}.}
\label{v2-Snn}
\end{figure}

In Fig.~\ref{v2-Snn}, the collision energy dependence of the $p_T$-integrated $v_2$ difference between mid-pseudorapidity $\pi^+$ and $\pi^-$ from the extended AMPT model is compared with those measured by the STAR Collaboration. At each collision energy, tens of thousands events were generated from transport simulations. The decreasing trend of $v_2$ splitting with increasing collision energy is qualitatively reproduced by the extended AMPT model. Although the mean values of the $v_2$ difference between $\pi^+$ and $\pi^-$ have a smaller magnitude compared with the experimental data, they have overlaps within statistical errors. Our calculations show that the $v_2$ difference between $\pi^+$ and $\pi^-$ at RHIC-BES energies favors a strong vector-isovector interaction of $R_{IV}>2$. It is worthy to point out that the strong vector-isovector coupling obtained in the present study does not affect the $v_2$ splitting between nucleons and antinucleons as well as between $K^+$ and $K^-$ obtained in Ref.~\cite{Xu14} since they are due to the vector-isoscalar coupling.

To summarize, we have studied the effect of the vector-isovector interaction on the elliptic flow difference between $\pi^+$ and $\pi^-$ based on the framework of an extended multiphase transport model, in which the partonic mean-field potentials are obtained from a 3-flavor Nambu-Jona-Lasinio model including isovector couplings, and the hadronization algorithm is improved by considering the coalescence of partons close in phase space. We have found that the time component of the vector-isovector potential leads to a more repulsive (attractive) potential for $d$ ($u$) quarks in the baryon-rich and $d$-quark-rich medium, thus enhancing the $v_2$ of $d$ quarks and $\pi^-$ while reducing that of $u$ quarks and $\pi^+$. The space component of the vector-isovector potential and the hadronic potentials are, however, found to have less important effect on the isospin splitting of $v_2$. With increasing strength of the vector-isovector coupling, the $v_2$ difference between $\pi^+$ and $\pi^-$ tends to saturate. Results from our transport approach reproduce the decreasing $v_2$ splitting with increasing collision energy seen in experiments, and the experimentally observed $v_2$ difference between $\pi^+$ and $\pi^-$ favors a strong vector-isovector coupling, with its coupling strength larger than twice the scalar-isoscalar coupling strength. In the $d$-quark-rich matter, a stronger vector-isovector coupling disfavors the splitting of $u$ and $d$ quark chiral phase transition boundary, leads to a critical point at higher temperatures, and results in a larger quark matter symmetry energy~\cite{Liu16}.

Although the total isospin splitting of pion $v_2$ can be explained by a strong vector-isovector interaction, its dependence on the charge asymmetry~\cite{STAR15} needs further investigations. Also, the isospin-independent polyakov-loop potential, which is essential for describing the QCD deconfinement phase transition and the thermodynamic properties of QGP~\cite{Liu16}, needs to be incorporated in transport simulations. These studies are in progress.

We thank Chen Zhong for maintaining the high-quality performance of the computer facility. J.X. acknowledges support from the Major State Basic Research Development Program (973 Program) of China under Contract No. 2015CB856904 and the National Natural Science Foundation of China under Grant Nos. 11922514 and 11421505. C.M.K. acknowledges support from the US Department of Energy under Contract No. DE-SC0015266 and the Welch Foundation under Grant No. A-1358.

\end{document}